\documentclass[prl,floatfix,twocolumn,footinbib]{revtex4}
\usepackage{epsfig}
\usepackage{amsmath,amssymb}
\begin{document}

\title{Self-sustained magnetoelectric oscillations in magnetic resonant tunneling structures}

\author{Christian Ertler\footnote{email: christian.ertler@physik.uni-regensburg.de}}
\author{Jaroslav Fabian}
\affiliation{Institute for Theoretical Physics, University of
Regensburg, Universit\"atsstrasse 31, D-93040 Regensburg, Germany}

\begin{abstract}

The dynamic interplay of transport, electrostatic, and magnetic
effects in the resonant tunneling through ferromagnetic quantum
wells is theoretically investigated. It is shown that the
carrier-mediated magnetic order in the ferromagnetic region not only
induces, but also takes part in intrinsic, robust, and sustainable
high-frequency current oscillations over a large window of nominally
steady bias voltages. This phenomenon could spawn a new class of
quantum electronic devices based on ferromagnetic semiconductors.

\end{abstract}

\maketitle

Ferromagnetism of diluted magnetic semiconductors (DMSs), such as
GaMnAs \cite{Ohno1998:S}, depends strongly on the carrier density
\cite{Dietl1997:PRB, Dietl:2007, Jungwirth2006:RMP, Lee2002:SST,
Jungwirth1999:PRB}. The possibility to tailor space charges in
semiconductors by bias or gate fields naturally suggests similar
tailoring of magnetic properties of DMSs. While early experiments
have indeed succeeded in generating ferromagnetism in DMSs
electrically or optically \cite{Ohno2000:N, Boukari2002:PRL},
ramifications of the strong carrier-mediated ferromagnetism in the
transport through DMS heterostructures are largely unexplored.

In resonant tunneling through a quantum well not only the tunneling
current, but also the carrier density in the well are sensitive to
the alignment of the electronic spectra in the leads and in the
well. If the quantum well is a paramagnetic DMS, the resulting
transport is influenced by the spin splitting of the carrier bands
in the well, as observed experimentally \cite{Slobodskyy2003:PRL}.
The magnetic resonant diodes are prominent spintronic devices
\cite{Fabian2007:APS, Zutic2004:RMP}, proposed for spin valves and
spin filtering \cite{Petukhov2002:PRL, Ertler2006a:APL}, or for
digital magnetoresistance \cite{Ertler2006b:APL,Ertler2007a:PRB}.

If the quantum well is made of a \emph{ferromagnetic} DMS
\cite{Oiwa2004:JMMM, Ohya2007:PRB}, resonant tunneling conditions
should influence magnetic ordering as well. It has already been
predicted that the critical temperature $T_c$ of the well can be
strongly modified {\em electrically} \cite{Ganguly2005:PRB,
Fernandez2002:PRB,Lee2000:PRB, Lee2002:SST}. Here we show that the
magnetic ordering affects back the tunneling current, in a peculiar
feedback process, leading to interesting dynamic transport
phenomena.

Conventional {\em nonmagnetic} RTDs can exhibit subtle intrinsic
bistability and terahertz current oscillations
\cite{Ricco1984:PRB,Jona-Lasinio1992:PRL, Orellana1997:PRL,
Jensen1991:PRL,Zhao2003:PRB} resulting from the nonlinear feedback
of the stored charge in the quantum well. Interesting phenomena
occur also in multiple quantum wells and superlattices, in which
electric field domains form whose dynamics leads to current
oscillations in the kHz-GHz range \cite{Bonilla2005:RPP}. This
effect has been exploited for spin-dependent transport by
incorporating paramagnetic quantum wells
\cite{Sanchez2001:PRB,Bejar2003:PRB}.

In this article we introduce a realistic model of a
self-consistently coupled transport, charge, {\em and} magnetic
dynamics and apply it to generic asymmetric resonant diodes with a
ferromagnetic quantum well to predict self-sustained, stable
high-frequency oscillations of the electric current and quantum well
magnetization. We formulate a qualitative explanation for the
appearance of these magnetoelectric oscillations. In essence,
ferromagnetic quantum wells exhibit strong nonlinear feedback to the
electric transport since the ferromagnetic order, which gives rise
to the exchange splitting of the quantum well subbands, is itself
mediated by the itinerant carriers. This together with the Coulomb
interaction which effectively modifies the single electron
electrostatic potential in the well, leads to a strong coupling of
the transport, electric, as well as magnetic properties of
ferromagnetic resonant tunneling structures.

\begin{figure}
 \centerline{\psfig{file=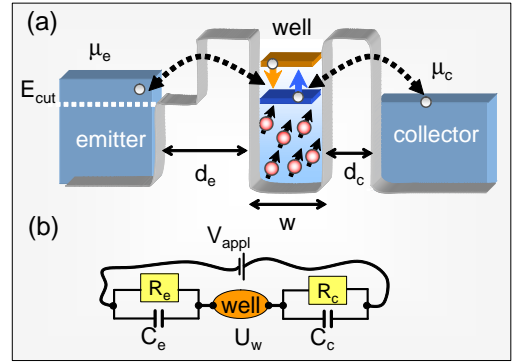,width=0.8\linewidth}}

\caption{(Color online) (a) Schematic scheme of the band profile of
the magnetic double barrier structure. The exchange interaction of
the magnetic ions is mediated by the carriers tunneling in and out
of the well. Here, the cut off of the emitter tunneling rate is
realized by a cascaded left barrier. (b) Equivalent circuit model of
the resonant tunneling structure introducing the emitter and
collector capacitances $C_e, C_c$ and resistances $R_e, R_c$,
respectively.} \label{fig:structure}
\end{figure}

Our model resonant tunneling structure with a ferromagnetic well
made of a DMS-material, e.g., GaMnAs, is sketched in
Fig.~\ref{fig:structure}(a). To exhibit magnetoelectric oscillations
the structure needs a built-in energy cut-off, $E_\mathrm{cut}$, of
the emitter tunneling rate. Such an energy cut-off might be realized
by a cascaded left barrier, as shown in Fig.~\ref{fig:structure}(a), by
which the tunneling for carriers with energies smaller than
$E_\mathrm{cut}$ is exponentially suppressed, due to the increased
barrier width. Other possibilities to realize the cutoff are
discussed below.

As we aim to understand the most robust features of the
ferromagnetic resonant tunnel structures, we present a minimal
theoretical model which captures the essential physics. The
longitudinal transport through the system can be described in terms
of sequential tunneling, as the high densities of magnetic
impurities residing in the ferromagnetic well will likely cause
decoherence of the propagating carries. Based on the transfer
Hamiltonian formalism a master equation for the semiclassical
particle distribution in the well can be derived, as described
elsewhere \cite{Fabian2007:APS, Averin1991:PRB}. We assume that
there is only a single resonant well level $E_0$ in the energy range
of interest, allowing us to write the rate equations for the
spin-resolved, time dependent quantum well particle densities
$n_\sigma(t), (\sigma = \uparrow,\downarrow=\pm 1/2)$ as,
\begin{eqnarray}\label{eq:rate}
\frac{\mathrm{d}n_\sigma}{\mathrm{d} t} &=& \Gamma_{e}(E_\sigma)\:
n_{e,\sigma} + \Gamma_{c}(E_\sigma)\:
n_{c,\sigma} - \Gamma(E_\sigma)\: n_\sigma\nonumber\\
&&-\frac{n_\sigma-n_{0,\sigma}}{\tau_s}.
\end{eqnarray}
Here,  $\Gamma_{\{e,c\}}$ denotes the energy-dependent tunneling
rate from the emitter (e) and the collector (c), $\Gamma =
\Gamma_e+\Gamma_c$ denotes the total tunneling rate, $\tau_s$ stands
for the spin relaxation time in the well, $n_{0,\sigma}$ denotes the
quasiequilibrium particle spin density, and $n_{\sigma,\{e,c\}}$ are
the densities of particles in the emitter and collector reservoir,
respectively, having the resonant longitudinal energy $E_\sigma$.
The spin-split resonant energies are
\begin{equation}\label{eq:ewell}
E_\sigma = E_0+U_w-\sigma\Delta,
\end{equation}
with $U_w$ being the electrostatic well potential and $\Delta$
denoting the subband exchange splitting. The total energy of the
particles $E_\mathrm{tot}$  is then given by the sum of the
longitudinal energy $E_\sigma$ and the in-plane kinetic energy:
$E_\mathrm{tot}= E_\sigma+T_{\mathrm{in}}$. The physical meaning of
the right side of Eq.~(\ref{eq:rate}) is as follows: the first two
terms are the gain terms, describing tunneling from the emitter and
the collector into the well; the third term describes all loss
processes due to the tunneling out of the well, and the last term
models the spin relaxation in the well. Considering the Fermi-Dirac
distributions in the emitter and the collector, the particle
densities $n_{i,\sigma}$ are,
\begin{equation}
n_{i,\sigma} = D_0 k_B T \ln\left\{1+\exp\left[(\mu_i-E_\sigma)/k_B
T\right]\right\}\quad i=e,c,
\end{equation}
with $k_B$ denoting Boltzmanns' constant; $T$ is the lead
temperature, $\mu_i$ are the emitter and collector chemical
potentials with $\mu_c = \mu_e-e V_\mathrm{appl}$ where
$V_\mathrm{appl}$ is the applied bias, and $D_0 = m/2\pi\hbar^2 $ is
the two-dimensional density of states per spin for carriers with the
effective mass $m$. The tunneling rates are essentially given by the
overlap of the lead and well wave functions according to Bardeen's
formula \cite{Bardeen1961:PRL}. For high barriers the rate becomes
proportional to the longitudinal momentum $p_z$ of the particles
\cite{Averin1991:PRB}, i.e., $\Gamma_{e,c} \propto (E_z)^{1/2}$ with
$E_z$ denoting the longitudinal energy.

In the framework of a mean field model for the carrier mediated
ferromagnetism in heterostructure systems
\cite{Dietl1997:PRB,Jungwirth1999:PRB, Lee2002:SST, Fabian2007:APS}
the steady state exchange splitting of the well subbands is
determined by
\begin{eqnarray}
\Delta_0 &=& J_\mathrm{pd}\int\mathrm{d}z\:
n_\mathrm{imp}(z)\left|\psi_0(z)\right|^2\nonumber\\
\label{eq:exchange}&& \times S B_S\left[\frac{S J_\mathrm{pd} s
(n_\downarrow- n_\uparrow) \left|\psi_0(z)\right|^2}{k_B T}\right].
\end{eqnarray}
Here, $J_\mathrm{pd}$ denotes the coupling strength between the
impurity spin and the carrier spin density (in case of GaMnAs p-like
holes couple to the d-like impurity electrons), $z$ is the
longitudinal (growth) direction of the structure,
$n_\mathrm{imp}(z)$ is the impurity density profile, $\psi_0(z)$
labels the well wave function, $B_S$ denotes the Brillouin function
of order $S$, and $S $ and $s= 1/2$ are the impurity and particle
spin, respectively. (In the case of Mn impurities S = 5/2.) The
expression shows that the well spin-splitting depends basically on
the particle spin polarization $\xi = n_\downarrow- n_\uparrow$ in
the well and the overlap between the wave function and the impurity
band profile. For simplicity, we consider here a homogenous impurity
distribution in the well, which makes $\xi$ the determining factor
for $\Delta_0$.

After a sudden change of the well spin polarization the magnetic
impurities need some time to respond until the corresponding mean
field value $\Delta_0$ is established. In the case of GaMnAs,
experimental studies of the magnetization dynamics revealed typical
response times of about 100 ps \cite{Wang2007:PRL}. We model the
magnetization evolution within the relaxation time approximation,
$\mathrm{d}\Delta/\mathrm{d}t = -(\Delta-\Delta_0)/\tau_\Delta$,
with $\tau_\Delta$ denoting the well spin-splitting relaxation time.

Finally, to take into account the nonlinear feedback of the Coulomb
interaction of the well charges, we introduce emitter-well and
collector-well capacitances, $C = C_e+C_c$ according to the
equivalent circuit model of a resonant tunneling diode, as shown in
Fig.~\ref{fig:structure}(b). The capacitances $C_e$ and $C_c$ are
determined by the geometrical dimensions of the barriers and the
well \cite{Averin1991:PRB}. The electrostatic well potential can
then be written as
\begin{equation}\label{eq:upot}
U_w = \frac{1}{C}\left[e^2(n-n_\mathrm{back})-C_c
eV_\mathrm{appl}\right]
\end{equation}
with $e$ denoting the elementary charge, $n_\mathrm{back}$ is the
positive background charge (from magnetic donors) in the well. All
the equations are nonlinearly coupled via Eq.~(\ref{eq:ewell}) for
the resonant well levels $E_\sigma$, making a numerical solution
indispensable.


\begin{figure}
 \centerline{\psfig{file=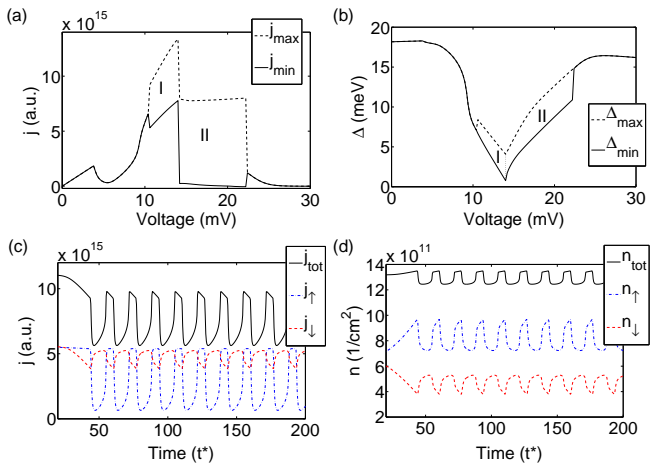,width=1\linewidth}}

\caption{(Color online) (a) Current-voltage (IV)-characteristic of
the investigated structure at the energy cut-off $E_\mathrm{cut} =
85$ meV. In the unsteady region between 11 to 22 mV the current does
not reach a steady state value; instead intrinsic current
oscillations appear, in which both the maximum and minimum values of
the oscillations are indicated by the dashed and solid lines,
respectively. Two different dynamic modes I and II can be
identified. (b) Well splitting $\Delta$ versus applied voltage. In
the unsteady region the maximum (dashed) and minimum (solid) values
of the magnetization oscillations are plotted. (c) and (d)
Transients of the spin-resolved current $j$ and the well particle
density $n$ at the applied voltage $V = 12$ mV and $E_\mathrm{cut} =
85$ meV. The time is measured in units of $t^*$ (typically some
picoseconds) as explained in the text.} \label{fig:IV}
\end{figure}

In the numerical simulations we use generic parameters assuming a
GaMnAs well: $m = 0.5\:m_0$, $\varepsilon_r = 12.9$, $d_e = 50$ \AA,
$d_c = 20$ \AA, $w = 10$ \AA, $\mu_e = 100$ meV, $n_\mathrm{imp} =
1.5\times10^{20} $cm$^{-3}$, $J_{\mathrm{pd}} = 0.06$ eV nm$^3$,
$\tau_s = 1\:t^*$, $\tau_\Delta = 10\:t^*$, where $d_e, d_c$ and $w$
are the emitter barrier, collector barrier and quantum well widths,
$m_0$ denotes the free electron mass, and $\varepsilon_r$ is the
relative permittivity of the well. The characteristic time scale
$t^*$ is the inverse emitter tunneling rate at the emitters' Fermi
energy, $[t^* = 1/\Gamma_e(\mu_e)]$, being of the order of
picoseconds. For the well background charge we consider that in
GaMnAs the actual carrier density is only about 10\% of the nominal
Mn doping density \cite{DasSarma2003:PRB}: $n_\mathrm{back} =
0.1\:n_\mathrm{imp}$. Our calculations are performed at $T = 4.2$ K,
which is well below the critical temperature of our specific well
$T_c \approx 10$ K, where we estimated $T_c$ by using the mean-field
result given in Ref.~\cite{Lee2000:PRB}.

Figure~\ref{fig:IV}(a) shows the current-voltage (IV) characteristic
of the structure. Up to about V = 11 mV the typical peaked IV-curve
of a resonant tunneling diode is obtained. However, in the voltage
range of 11 to 22 meV, which we will call hereafter the ``unsteady''
region, the current does not settle down to a steady state value;
instead stable high-frequency oscillations occur, as shown in
Fig.~\ref{fig:IV}(c) for the applied voltage of $V = 12$ mV. The
current is always evaluated at the collector side: $j_{c,\sigma}  =
\Gamma_c(E_\sigma)(n_\sigma-n_{c,\sigma})$. Along with the current,
oscillations of the well magnetization as well as of the spin
densities appear, as shown in Fig~\ref{fig:IV}(b) and (d). Those
magneto-electric oscillations are the main results of this paper.

\begin{figure}
\centerline{\psfig{file=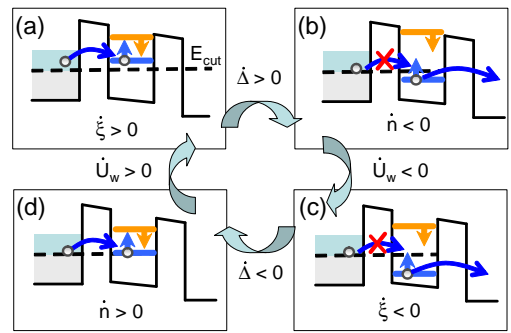,width=0.8\linewidth}}

\caption{(Color online) Explanation for the occurrence of
self-sustained oscillations in the case of mode I. For mode II the
arguments are completely analogous but there the spin down level is
crossing the cut-off energy $E_\mathrm{cut}$. (a) According to the
in-tunneling spin-up carriers the spin polarization $\xi$ is
increased, giving rise to an increasing well splitting $\Delta$. (b)
When the spin up level falls below the cutoff energy the total
particle number decreases and, hence, also the electrostatic
potential $U_w$ does. (c) This pushes the spin-up level even deeper
into the cut-off region leading to a fast decrease of the spin
polarization and consequently of the well splitting, which brings
the spin up level back to the emitter's supply region (d),
restarting the whole cycle.} \label{fig:scheme}
\end{figure}

The IV-curve in the unsteady region suggests the existence of two
qualitatively different dynamic modes. Indeed, comparing the
transients in these two voltage regions reveals that in region I the
spin up level is recurringly crossing the emitter energy cut off
$E_\mathrm{cut}$, whereas in region II this is done by the spin down
level, as schematically illustrated in Fig.~\ref{fig:scheme}. This
insight offers the following explanation for the the occurrence of
self-sustained oscillations. Take mode I; the arguments for mode II
are similar. The dropping of the spin up level below the cut-off
energy (due to the increasing exchange splitting) as two
implications: (i) the supply of the emitter spin up electrons
sharply decreases. Hence, the total well particle density $n =
n_\uparrow+ n_\downarrow$ decreases because the spin up electrons
residing in the well are tunneling out to the collector. A decreased
particle density leads to a decreased electrostatic potential
according to Eq.~(\ref{eq:upot}), which effectively drives the spin
up level even deeper into the cut-off region. (ii) Since the spin up
electrons are the majority spins in the well, a decrease of
$n_\uparrow$ implies a decreasing spin polarization $\xi$ in the
well. This causes, via Eq.~(\ref{eq:exchange}), a rapid decrease of
the subband exchange splitting, bringing the spin up level back to
the emitter supply region. The spin up electrons can then tunnel
again into the well and the whole process starts from the beginning,
producing the calculated cycles.

The occurrence of these oscillations needs the concurrent interplay
of \emph{both} the electric and magnetic feedbacks: the
electrostatic feedback acts like an ``inertia'' for the
oscillations, allowing the spin level to get deeper into the cut-off
region, whereas the magnetic feedback is needed to bring the level
back into the emitter's supply region. From the above discussions it
also follows that a steep descent in the tunneling rate at the
cut-off energy is necessary. This is confirmed in our simulations,
where we assumed an exponential decay of $\Gamma_E$ for $E_z <
E_\mathrm{cut}$. Actually, this is also the reason why no
oscillations are occurring at the emitter's conduction band edge
$(E_z = 0)$, which is an otherwise natural emitter's cut-off,
because near the band edge the tunneling rate is already almost
vanishing according to $\Gamma_{E} \propto (E_z)^{1/2}$.

\begin{figure}
 \centerline{\psfig{file=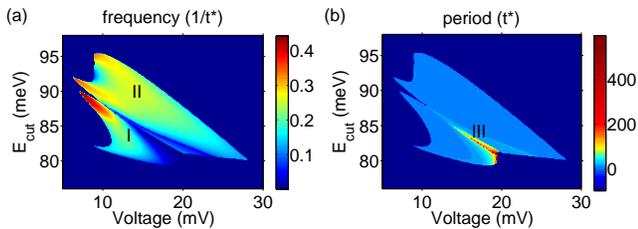,width=1\linewidth}}

\caption{(Color online) Contour plots of the frequency (a) and the
period (b) of the intrinsic oscillations as a function of the
applied voltage and the cut-off energy $E_\mathrm{cut}$. The two
islands of high-frequency oscillations correspond to the dynamic
modes I and II. They are separated by a crossover region III of
high-period oscillations, which becomes most evident in period
contour plot (b). } \label{fig:phdiag}
\end{figure}

The typical timescale for the oscillations is made up by two
contributions: (i) the evolution of the well splitting $\Delta$ in
the emitter's supply region, i.e., the time needed for the well
splitting $\Delta$ to become large enough that one of the spin
levels crosses the cut off energy, and (ii) the dynamics of $\Delta$
in the cut-off region. The timescale of contribution (i) can be
estimated to be proportional to $\tau_\Delta(r-1)$, where $r$ is the
ratio of the well splitting when the cut-off energy is reached to
the initial splitting at the beginning of the cycle, whereas the
dynamics of contribution (ii) is mostly governed by a fast
rearrangement of the spin densities in the well, which happens on
the order of the tunneling time $t^*$. This is in line with the
numerical result that the oscillation frequency increases with
decreasing $\tau_\Delta$. In contrast, a decreasing spin relaxation
time, which diminishes more and more effectively the spin
polarization in the well and, hence, the well splitting, gives rise
to a decrease of the frequency.

The fastest magnetoelectric oscillations in our simulations for the
generic parameters used exhibit periods of about $2.5\times
t^*\approx 5$ ps. Important, the oscillation frequency can be
modified by varying the applied voltage, as can be seen in
Fig.~\ref{fig:phdiag}, which displays contour plots of the frequency
and the period versus the applied voltage and the cut-off energy.
Two ``islands'' can be distinguished, corresponding to the two
regions of the dynamic modes I and II, respectively (see
Fig.~\ref{fig:IV}). Mode II appears at higher voltages, where the
collector chemical potential is already below the cut-off energy. In
this case the decrease of the well particle density is much stronger
when the spin up level crosses the cut-off energy, as illustrated in
Fig.~\ref{fig:scheme}(b), since none of the out-tunneling electrons
are ``Pauli-blocked" by the collector electrons. This gives rise to
a much stronger decrease of the electrostatic potential as compared
to mode I, pushing {\em both} the spin up and down levels below the
cut-off energy. In the following oscillations this makes the spin
down level recurringly crossing the cut-off energy instead of the
spin up level as is the case for mode I.

Both modes are separated by a crossover region III, in which the
oscillations have much longer periods than on the ``islands", as can
be seen in Fig.~\ref{fig:phdiag}(b). The reason for these
low-frequency oscillations is that the initial spin polarization
and, hence, the well splitting at the beginning of each cycle is
small, yielding a large ratio $r$. Therefore, in region III it can
take 10-100 times longer that the well splitting reaches the cut-off
energy as for modes I or II. The possibility of controlling the
frequency of the magnetoelectric oscillations by electrical means is
especially interesting from the applications point of view.

For the experimental observation of the self-sustained oscillations
the emitter needs to emit electrons in a sharply defined energy
window. In the model above we propose an asymmetric barrier, as in
Fig.~\ref{fig:structure}(a). Another practical way would be using an
auxiliary resonant tunneling structure in the emitter itself,
providing a spectral filter allowing only resonant electrons to pass
through. Yet another possibility would be employing a hot electron
emitter, say using a Schottky barrier.

We have shown that the charge and magnetization dynamics in
ferromagnetic tunneling  heterostructures are influenced by the
highly nonlinear feedback of {\em both} Coulomb and magnetic
couplings on the tunneling transport. The feedback results in
high-frequency self-sustained intrinsic oscillations suggesting
applications of ferromagnetic QWs in tunable high-power current
oscillators.

This work has been supported by the Deutsche Forschungsgemeinschaft
SFB 689.

\bibliography{spin}

\end{document}